\documentclass[aps,prd,twocolumn,aps,floatfix,showpacs,tightenlines,showkeys,superscriptaddress,amsmath,amssymb,nofootinbib]{revtex4-1}
\usepackage{amssymb,amsbsy,epsfig,color,graphicx}
\usepackage{color}
\usepackage{[longtable}
\usepackage{array}
\usepackage{dcolumn}   % needed for some tables
\usepackage{cellspace}
\usepackage{mathtools}
\usepackage{amstext}
\usepackage{amssymb}
\usepackage{stmaryrd}
\usepackage{stackrel}
\usepackage{graphicx}
\usepackage{esint}
\usepackage[utf8]{inputenc}
\usepackage{blindtext}
\usepackage{float}
\restylefloat{table}
\usepackage{booktabs}
\usepackage{enumitem} 
\usepackage{xcolor}
\usepackage{etoolbox} % for \appto
\usepackage{lipsum} % for mock text
\usepackage[capitalize]{cleveref}

\usepackage{multirow}
\usepackage[caption=false]{subfig}
\renewcommand\[{\begin{equation}}
\renewcommand\]{\end{equation}}

\newcommand{\ba}{\begin{eqnarray}}
\newcommand{\ea}{\end{eqnarray}}

\makeatletter

\appto{\appendix}{%
\@ifstar{\def\theequation@prefix{A.}}%
{}%
}
\makeatother

\begin{document}

\title{Non-singular metric for an electrically charged point-source in ghost-free infinite derivative gravity}

\author{Luca Buoninfante}
\affiliation{Dipartimento di Fisica "E.R. Caianiello", Universit\`a di Salerno, I-84084 Fisciano (SA), Italy}
\affiliation{INFN - Sezione di Napoli, Gruppo collegato di Salerno, I-84084 Fisciano (SA), Italy}
\affiliation{Van Swinderen Institute, University of Groningen, 9747 AG, Groningen, The Netherlands}
\author{Gerhard Harmsen}
\affiliation{National Institute for Theoretical Physics; School of Physics, University of the Witwatersrand, Johannesburg, Wits 2050, South Africa.}
\affiliation{Van Swinderen Institute, University of Groningen, 9747 AG, Groningen, The Netherlands} 

\author{Shubham Maheshwari}
\affiliation{Van Swinderen Institute, University of Groningen, 9747 AG, Groningen, The Netherlands}

\author{Anupam Mazumdar}
\affiliation{Van Swinderen Institute, University of Groningen, 9747 AG, Groningen, The Netherlands}
\affiliation{Kapteyn Astronomical Institute, University of Groningen, 9700 AV, Groningen, The Netherlands}

%	\date{\today}

\begin{abstract}
In this paper we will construct a linearized metric solution for an electrically charged system in a {\it ghost-free} infinite derivative theory of gravity which is valid in the entire region of spacetime. We will show that the gravitational potential for a point-charge with mass $m$ is non-singular, the Kretschmann scalar is finite, and the metric approaches conformal-flatness in the ultraviolet regime where the non-local gravitational interaction becomes  important. We will show that the metric potentials are bounded below one as long as two conditions involving the mass and the electric charge are satisfied. Furthermore, we will argue that the cosmic censorship conjecture is not required in this case. Unlike in the case of Reissner-Nordstr\"om in general relativity, where $|Q|\leq m/M_p$ has to be always satisfied, in {\it ghost-free}  infinite derivative gravity  $|Q|>m/M_p$ is also allowed, such as for an electron.

\end{abstract}

\maketitle

%%%%%%%%%%%%%%%%%%%%%%%%%%%%%%%%%%%%%%%%%%%%%%%%%%%%%%%%%%%%%%%%%%%%%%%%

\section{Introduction}\label{intro}

Einstein's theory of general relativity (GR) has been tested to a very high precision in the infrared (IR) regime, i.e. at large distances and late times \cite{-C.-M.}; including  the recent detections of gravitational waves due to merging of binaries which matches excellently the predictions of GR~\cite{-B.-P.}. However, as soon as one approaches short distances, the theory shows several problems, as for example black-hole and cosmological singularities at the classical level, and fails to be perturbatively renormalizable at the quantum level. Indeed, GR turns out to be incomplete in the ultraviolet (UV) regime, i.e. at short distances and small time scales.

It is well known that by adding quadratic terms in the curvature to the Einstein-Hilbert action one obtains a gravitational theory which becomes power counting renormalizable~\cite{-K.-S.}. In the UV, the theory becomes conformal and in the 
infrared Einstein's GR is recovered, except that being a higher derivative theory of gravity, it suffers from instability due to the presence of a massive spin-$2$ {\it ghost}-like degree of freedom in the physical spectrum, whose emergence becomes evident by computing the graviton propagator around the Minkowski background \cite{-K.-S.}.

Recently, it has been shown that the {\it infinite derivative gravity} (IDG) can resolve this {\it ghost } problem and, at the same time, it can also resolve the short-distance singularity for a point-like object, regularizing the $1/r$ behavior of the metric potential in the linearized limit, therefore ameliorating the UV aspects of Einstein's and Newtonian gravity~\cite{Biswas:2005qr,Biswas:2011ar}~\footnote{Prior to Ref.~\cite{Biswas:2011ar}, there have been 
discussions on resolving singularities in infinite derivative gravity in Refs.~\cite{-Yu.-V.,Tomboulis,Tseytlin:1995uq,Siegel:2003vt}.}. 
Furthermore, the time-dependent case yields a vacuum solution which is devoid of any cosmological singularity \cite{Biswas:2005qr,Koivisto,Koshelev:2012qn}. It was also shown that the theory has a mass-gap at the linear level~\cite{Frolov},
and never develops a singularity even in a dynamical setup~\cite{Frolov:2015bia,Frolov:2015usa}. At the linearized level, the non-trivial gravitational solution, in the UV, gives rise to constant Ricci scalar, Ricci tensor and Riemann tensor, and the Weyl tensor approaches zero at the origin (quadratically in distance from  $r=0$ at best). The Kretschmann scalar remains constant and never blows up at the origin~\cite{Buoninfante:2018xiw}. Also, the static solution has now been extended to a non-singular rotating metric which resolves the Kerr-type ring singularity in Einstein's GR~\cite{Cornell:2017irh}. In Ref.~\cite{Boos:2018bxf}, then authors have studied gravitational potentials in extended objects within {\it ghost-free} IDG, and found to be free from singularity.

Very recently, it has been shown that the full non-linear equations of motion for a ghost-free infinite derivative gravity, given by~\cite{Biswas:2013cha}, does not provide a  $1/r^{\alpha}$, with $\alpha >0$, as a full vacuum solution~\cite{Koshelev:2018hpt}, where the Weyl term played the crucial role. Furthermore, in the time dependent context the infinite derivative gravity does not give rise to the Kasner-type metrics as vacuum solutions, and therefore provides a way to avoid the BKL-singularity which describes an anisotropic collapse of a time dependent metric solution in the case of the Einstein-Hilbert action \cite{Koshelev:2018rau}. 

As pointed out in Ref. \cite{conformal}, the usual notion of {\it vacuum} solution that we use in GR, for example for the Schwarzschild solution, does not apply to the case of infinite derivative gravity. Indeed, the delta source distribution at $r=0$, which is imposed as a boundary condition, is smeared out by the infinitely many derivatives, generating an extended source whose size is given by the scale of non-locality, $r_{NL}\sim 2/M_s$ \cite{Koshelev:2017bxd}. This point is crucial in order to obtain a full metric solution which is regular at $r=0.$ 

Infinite derivative gravity has also been discussed in the context of field theory. It is expected that infinite derivatives will ameliorate the quantum aspects as well, and it has been argued that such a class of theory will be super-renormalizable, see~\cite{-Yu.-V.,Tomboulis,Talaganis:2014ida}. Moreover, high energy scattering of gravitons in this class of theory does not necessarily lead to a formation of a blackhole as shown in Ref.~\cite{Talaganis-scat}. The scattering amplitude decreases exponentially, if the propagator and vertex corrections are correctly taken into account~\cite{Talaganis:2014ida,Talaganis-scat}. Furthermore, in quantum field theory, infinite derivatives can be useful to understand the UV completion of the Standard Model~\cite{Biswas:2014yia}, and also the stability of the Higgs within the Abelian-Higgs model~\cite{Ghoshal:2017egr}.  

Inspite of all these interesting results, there are many important challenges and unanswered riddles regarding IDG and the non-local regime of gravity, such as renormalizability, concept of spacetime, unitarity, causality structure, etc., indeed some of these issues have been discussed partly in Refs.~\cite{Woodard,Deser:2007jk,Frolov:2017rjz,Chin:2018puw,Buoninfante:2018mre}, but a clear quantum picture would be more desirable, and goes beyond the scope of the current paper. 
Among other things, the formulation of the initial value problem within the context of infinite-derivatives theories remains an interesting point of discussion~\cite{Barnaby:2007ve}. 

The aim of this paper is to show how infinite derivative gravity will modify the gravitational potential generated by a charged source; we will produce a non-singular generalization of the Reissner-Nordstr\"om metric. We will briefly review the infinite derivative ghost-free gravity, then we will discuss the linearized metric of an electrically charged source. We will study various spacetime properties and discuss the consequences for the {\it cosmic censorship} hypothesis~\cite{penrose,hawking,wald}.

%%%%%%%%%%%%%%%%%%%%%%%%%%%%%%%%%%%%%%
\section{Infinite derivative ghost-free gravity}\label{idg}

The most general quadratic curvature gravity in $4$ dimensions, which is parity invariant and torsion free around a constant curvature background can be written 
as~\cite{Biswas:2011ar,Biswas:2016etb}
\begin{equation}\label{eq:1}
\begin{array}{rl}
S= & \displaystyle\frac{1}{16\pi G} \displaystyle\int d^{4}x\sqrt{-g}\left[{ \mathcal{R}}+\alpha\left(\mathcal{R}\mathcal{F}_{1}(\Box_{s})\mathcal{R}\right.\right.\\
& \displaystyle \left.\left.+\mathcal{R}_{\mu\nu}\mathcal{F}_{2}(\Box_{s})\mathcal{R}^{\mu\nu}+\mathcal{R}_{\mu\nu\rho\sigma}\mathcal{F}_{3}(\Box_{s})\mathcal{R}^{\mu\nu\rho\sigma}\right)\right], 
\end{array}
\end{equation}
where $G=1/M_p^2$ is Newton's constant and $\alpha\sim 1/M_s^2$ is a dimensionful coupling. We are working with mostly positive metric signature $(-,+,+,+)$, and $\Box_s \equiv \Box/M_s^2$, where the d'Alembertian is given by: 
$\Box= g^{\mu\nu}\nabla_{\mu}\nabla_{\nu}$, and $\mu,\nu=0,1,2,3$. The new scale of gravity is given by $M_s\leq M_p$~~\footnote{We work in Natural Units: $\hbar=1=c.$}. The three gravitational form factors ${\cal F}_{i}$ are reminiscent of any massless theory, and are constrained by the tree-level 
unitarity and {\it ghost-free} condition derived in Ref.~\cite{Biswas:2011ar}. Around the Minkowski background we can avoid the presence of other dynamical degrees of freedom other than the spin-$2$ massless graviton by first demanding 
\begin{equation}
2{\cal F}_1(\Box_s)+{\cal F}_2(\Box_s)+ 2{\cal F}_3(\Box_s)=0\,.
\end{equation}
Thus, the propagator of this theory is given by~\cite{Biswas:2011ar,Biswas:2013kla,Buoninfante}
\begin{equation}
\Pi(-k^2)=\frac{1}{a(-k^2)}\left(\frac{\mathcal{P}^{(2)}}{k^2} -\frac{\mathcal{P}_s^{(0)}}{2k^2}\right)\,,\label{propag}
\end{equation}
where $\Pi_{GR}=\mathcal{P}^{(2)}/k^2 -\mathcal{P}_s^{(0)}/2k^2$ is the graviton GR propagator
and $\mathcal{P}^{2},$ $\mathcal{P}_s^{0}$ are the two spin projection operators along the spin-$2$ and spin-$0$ components, respectively; while\footnote{Around Minkowski background, up to quadratic order in the metric perturbation, the form factor $\mathcal{F}_3(\Box_s)$ can be set to zero without any loss of generality.} 
\begin{eqnarray}
a(\Box_s)=1-\frac{1}{2}{\cal F}_2(\Box_s)\Box_s -2{\cal F}_3(\Box_s)\Box_s. \label{choice}
\end{eqnarray}
Moreover, in order to have {\it no} ghost-like degree of freedom, the form of $a(\Box)$ is restricted to be {\it exponential of an entire} function, so that {\it no} new poles are introduced in the propagator in Eq. \eqref{propag}.
One simple choice is indeed 
\begin{equation}\label{eq:2}
a(\Box_{s})=e^{-\Box/M_s^{2}}\,.
\end{equation}
Other choices of {\it entire function} have been explored in Refs.~\cite{Edholm:2016hbt}, which illustrates universal behavior in the Newtonian potential from IR all the way up to the UV regime.

In Ref. \cite{Biswas:2011ar} it was shown that the ghost-free condition, like in Eq. \eqref{eq:2}, also yields regular static solutions in the linear regime, where the metric potential satisfies: $2|\Phi|<1.$ Furthermore, in Ref. \cite{conformal}, it was argued that this inequality can always hold true when the quadratic part of the action in Eq. \eqref{eq:1} dominates over the Einstein-Hilbert term at the horizon scales~\footnote{Indeed, by considering a characteristic length scale $L,$ such that $dx\sim L,$ $\partial_x\sim 1/L$ and $\mathcal{R}\sim 1/L^2,$ and assuming $L\sim 2Gm,$ where $m$ is the mass of the object in the static geometry, we obtain $S\sim \frac{m^2}{M_p^2}\left(1+\frac{M_p^4}{m^2M_s^2}\right)$,
which tells us that the quadratic part of the action dominates over the Einstein-Hilbert term if and only if
$mM_s<M_p^2 $~\cite{conformal}. Moreover, as long as this condition holds, the static metric potential is always bounded by one.}.
In this paper we will show that a similar scenario also holds when we consider the spacetime metric for a charged point-source, but in this case an additional inequality, involving the electric charge, will be required in order to have bounded metric potentials.

%%%%%%%%%%%%%%%%%%%%%%%%%%%%%%%%%%%%%%%%%%%
\section{Linearized metric solution for an electrically charged source}\label{lin-metric}

We now want to determine the spacetime metric for an electrically charged source in infinite derivative gravity, in the linear regime. The linearized field equations corresponding to the action in Eq. \eqref{eq:1}, with the ghost-free choice in Eq. \eqref{eq:2} are given by \cite{Biswas:2011ar}:
\begin{equation}\label{eq:3}
\begin{array}{ll}
\displaystyle e^{-\Box/M_s^2}\left[\Box h_{\mu\nu}+(\eta_{\mu\nu}\partial_{\rho}\partial_{\sigma}h^{\rho\sigma}+\partial_{\mu}\partial_{\nu}h)\right. & \\
\,\,\,\,\,\,\,\,\,\,\,\displaystyle \left.-\partial_{\sigma}(\partial_{\nu} h^{\sigma}_{\mu}+\partial_{\mu}h_{\nu}^{\sigma})-\eta_{\mu\nu}\Box h\right]=-16\pi G\tau_{\mu\nu}. &
\end{array}
\end{equation}
The two-rank tensor $\tau_{\mu\nu}$ stands for the electro-magnetic energy-momentum tensor which is defined by:
\begin{equation}\label{eq:4}
\tau_{\mu\nu}=\frac{1}{4\pi}\left(\eta_{\rho\nu}F_{\mu\sigma}F^{\rho\sigma}-\frac{1}{4}\eta_{\mu\nu}F_{\rho\sigma}F^{\rho\sigma}\right),
\end{equation}
where $F_{\mu\nu}=\partial_{\mu}A_{\nu}-\partial_{\nu}A_{\mu}$ is the electro-magnetic field strength, with $A_{\mu}$ being the potential-vector. We will consider the case in which the source is static and spherically symmetric, and no magnetic monopole is present, thus the only non-vanishing contributions of the electromagnetic field-strength will come from:
\begin{equation}
F_{10}=-F_{01}= E_r \quad \text{and} \quad E_r = \frac{Q}{r^2}, \label{eq:4.1}
\end{equation}
where $E_r$ is the radial component of the electric field, with the Coulomb constant $k_e = 1$ in natural (Gaussian) units. We will work in the case of weak gravitational field and static source, so that the metric solution can be written as:
\begin{equation}
ds^2=-(1+2\Phi(r))dt^2+(1-2\Psi(r))(dr^2+r^2d\Omega^2),\label{eq:5}
\end{equation}
where $(t,r,\theta,\varphi)$ are the isotropic coordinates, $d\Omega^2=d\theta^2+{\rm sin}^2 \theta d\varphi^2,$ and $\Phi$ and $\Psi$ are the two unknown gravitational fields that need to be found. By looking at the $00$-component and trace of the linearized field equations in Eq. \eqref{eq:3} we obtain, respectively:
\begin{equation}
\begin{array}{rl}
\displaystyle e^{-\nabla^2/M_s^2}\left[\nabla^2h_{00}-\partial_i\partial_jh^{ij}+\nabla^2h\right]=&-16\pi G\tau_{00},  \\
\displaystyle e^{-\nabla^2/M_s^2}\left[-2\nabla^2h+2\partial_i\partial_jh^{ij}\right]=&-16\pi G\tau.
\end{array}\label{eq:6}
\end{equation}
From the metric-form in Eq. \eqref{eq:5} we note that $h=2(\Phi-3\Psi),$ $h_{00}=-2\Phi$ and $h_{ii}=-2\Psi$, thus we obtain two differential equations for the two metric potential $\Phi$ and $\Psi$:
\begin{equation}
\begin{array}{rl}
\displaystyle e^{-\nabla^2/M_s^2}\nabla^2\Phi= & \displaystyle 4\pi G(\tau+2\tau_{00}),  \\
\displaystyle e^{-\nabla^2/M_s^2}\nabla^2\Psi= & \displaystyle 4\pi G\tau_{00}.
\end{array}\label{eq:7}
\end{equation}
In the case of an electric charged static source the energy-momentum tensor in Eq. \eqref{eq:4} turns out to be traceless, $\tau=0,$ while the $00$-component is given by $\tau_{00}=Q^2/8\pi r^4,$ so that Eq. \eqref{eq:7} reads
\begin{equation}
\begin{array}{rl}
e^{-\nabla^2/M_s^2}\nabla^2\Phi=&\displaystyle \frac{GQ^2}{r^4},  \\
 e^{-\nabla^2/M_s^2}\nabla^2\Psi=&\displaystyle \frac{GQ^2}{2r^4}.
\end{array}\label{eq:8}
\end{equation}
The details of solving the differential equations above are laid out in Appendix \ref{general-sol}. Here, we state the final result for the two metric potentials:
\begin{equation}
\begin{array}{rl}
\displaystyle \Phi(r)=&\displaystyle-\frac{Gm}{r}{\rm Erf}\left(\frac{M_s r}{2}\right)+\frac{GQ^2M_s}{2r}{\rm F}\left(\frac{M_s r}{2}\right),  \\
\displaystyle \Psi(r)=&\displaystyle-\frac{Gm}{r}{\rm Erf}\left(\frac{M_s r}{2}\right)+\frac{GQ^2M_s}{4r}{\rm F}\left(\frac{M_s r}{2}\right),
\end{array}\label{eq:15}
\end{equation}
where 
\begin{equation}
{\rm Erf}(x):=\frac{2}{\sqrt{\pi}}\int\limits_0^x e^{-t^2}dt,
\end{equation}
is the Error function and 
\begin{equation}
{\rm F}(x):=e^{-x^2}\int\limits_0^x e^{t^2}dt, \label{dawson}
\end{equation}
\begin{figure}[t]
	\begin{center}
		\includegraphics[scale=0.47]{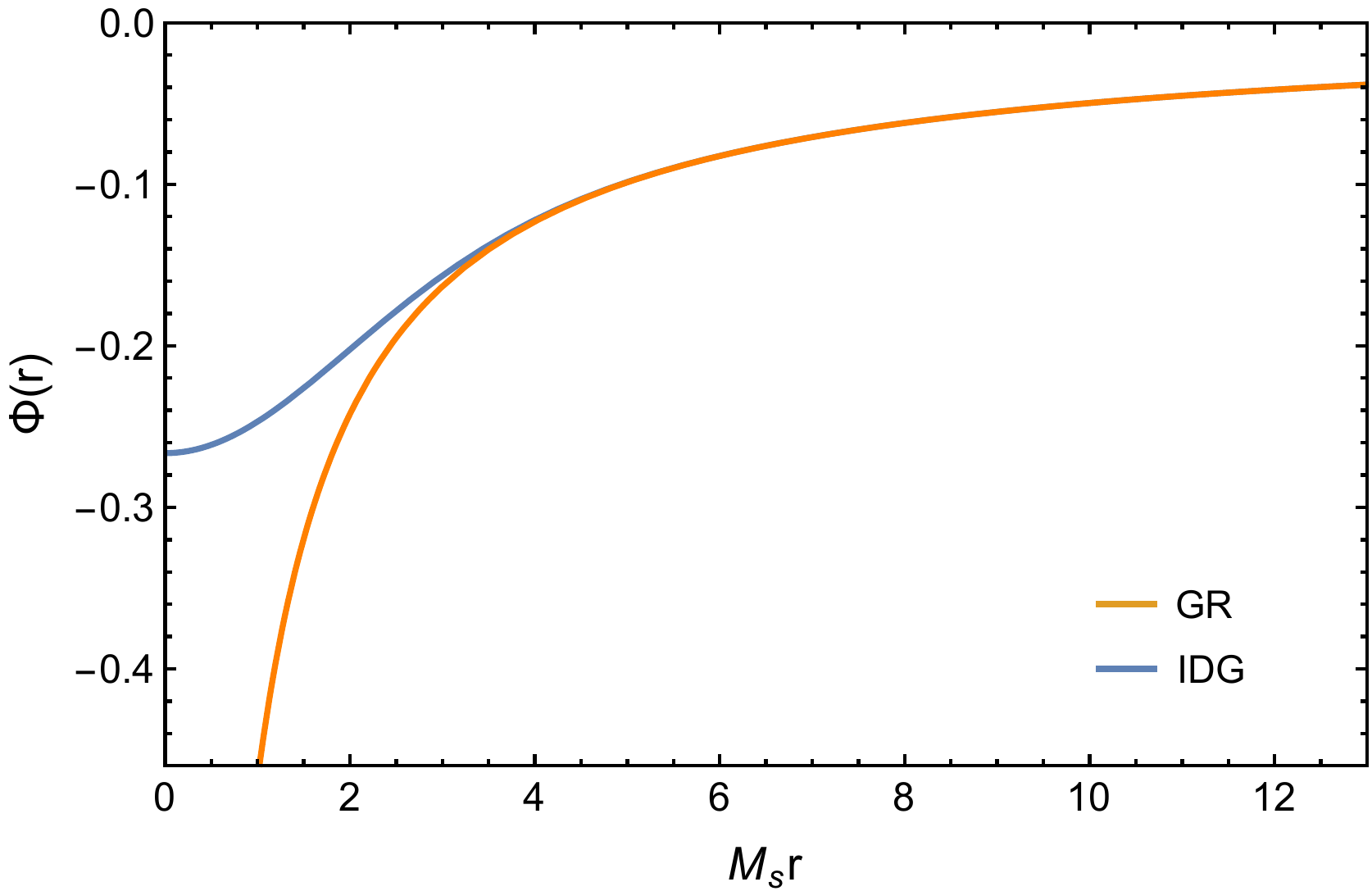}
		\caption{We have plotted the behavior of the potential $\Phi$ in IDG (blue line) and made a comparison with the corresponding one in the case of Reissner-Nordstr\"om metric in GR (orange line). We have chosen the values $G=1, m=1$ and $M_s = 0.5$ and take $Q=0.5$.}
	\end{center}\label{Fig:potential}
\end{figure}
is the Dawson function. Note that $\Phi\neq\Psi$, as it also happened in the GR case (see Appendix \ref{Reiss-Nord}). In the case $Q=0$ we recover the linearized IDG metric for a static neutral point-source derived in Ref. \cite{Biswas:2011ar}, as expected. In the IR regime, $M_s r \gg 2$, the potentials in Eq. \eqref{eq:15} reduce to those of the linearized Reissner-Nordstr\"om metric in GR, Eq. \eqref{ap:14}, as the asymptotic expansions for the Error function and the Dawson function are ${\rm Erf}(x)\sim 1$ and ${\rm F}(x) \sim 1/2x$.
More interestingly, in the non-local regime, $M_s r \ll 2$, the metric potentials $\Phi(r)$ and $\Psi(r)$ approach non-singular constant values given by:
\begin{equation}
\begin{array}{ll}
\lim\limits_{M_s r \to 0} \Phi(r) \equiv -A =\displaystyle-\frac{G m M_s}{\sqrt{\pi }}+\frac{1}{4} G Q^2 M_s^2,& \\
\lim\limits_{M_s r \to 0} \Psi(r) \equiv -B =\displaystyle-\frac{G m M_s}{\sqrt{\pi }}+\frac{1}{8} G Q^2 M_s^2,&
\end{array}\label{uv-region}
\end{equation}
because for $x \ll 1$ one has ${\rm Erf}(x) \sim 2x/\sqrt{\pi}$ and ${\rm F}(x) \sim x$. Note that when 
$Q=0$, we recover the case of neutral source, $A=B = GmM_s/\sqrt{\pi}$. 

In Fig. 1 the potential $\Phi$ has been plotted as a function of $M_sr$ and a comparison between {\it ghost-free} IDG and GR has been made. We have chosen the values $G=1, m=1,$ $M_s = 0.5$ and $Q=0.5$. It is also worth mentioning that the force per unit mass, $F_{g} (r) = -\partial \Phi/\partial r$, goes to zero linearly as a function of $r$. This vanishing force shows a classical aspect of asymptotic freedom in {\it ghost-free} IDG~\cite{Biswas:2011ar}.

Since we are working in the linearized regime, we need to satisfy the following two inequalities: $2|\Phi| < 1$ and $2|\Psi | < 1$. First of all, note that it is sufficient to focus just on one of the two inequalities, as the two metric potentials are almost the same apart from a factor $1/2$ appearing in the gravitational contribution due to the charge; see 
Eqs. (\ref{eq:15},\ref{uv-region}). Moreover, for simplicity, we can drop all numerical factors and just focus on the essential parameters $m,$ $G=1/M_p^2$, $M_s$ and $Q.$

Mathematically, the weak-field inequality can be also satisfied when the neutral and charged contributions to the gravitational potential are both very large but such that the modulus of their sum is still less than one. However, from a physical point of view, in order to satisfy the weak-field inequalities, we need to require that both contributions have to be smaller than one\footnote{As pointed out in Ref.\cite{conformal}, the inequality $mM_s< M_p^2$  always holds in order for the quadratic part of the gravitational action to dominate over the Einstein-Hilbert term in the UV regime. If the inequality was not satisfied, we would not be able to modify the Reissner-Nordstr\"om geometry and avoid the metric singularity at the origin.}~\cite{Biswas:2011ar}:  
\begin{equation}
mM_s<M_p^2, \label{1-inequality}
\end{equation}
\begin{equation}
Q^2M_s^2<M_p^2 \Longleftrightarrow |Q|M_s<M_p. \label{2-inequality}
\end{equation}
We can immediately notice that, with respect to the neutral case, we now have the additional inequality in Eq. \eqref{2-inequality}; together they will play a crucial role. In fact, we can make the following important observations.
\begin{itemize}
	
	\item  In the case of the Reissner-Nordstr\"om metric in Einstein's GR one has to demand that the inequality 
	\begin{equation}
	|Q|\leq \frac{m}{M_p} \label{1mass-charge-inequality} 
	\end{equation}
	always holds true in order to avoid the formation of a {\it naked singularity} during a collapsing phase. This is one example of the weak formulation of the {\it cosmic censorship conjecture}; see Refs. \cite{penrose,hawking,wald}. In the case of a charged source in {\it ghost-free} IDG there is {\it no} need to require a cosmic censorship conjecture as the theory turns out to be singularity-free and devoid of any horizon as long as the inequalities in Eqs. (\ref{1-inequality},\ref{2-inequality}) hold true. 
	
	\item Thus, in Einstein's GR one cannot describe the gravitational field generated by objects for which
    \begin{equation}
    |Q|> \frac{m}{M_p} \label{2mass-charge-inequality},
    \end{equation}
	such as for an electron. In {\it ghost-free} IDG, both inequalities in Eqs. (\ref{1mass-charge-inequality},\ref{2mass-charge-inequality}) are allowed. Indeed, consistently with the weak-field inequalities in Eqs. (\ref{1-inequality},\ref{2mass-charge-inequality}), we can have {\it two} possible scenarios:
	\begin{equation}
	 |Q|<\frac{m}{M_p}<\frac{M_p}{M_s} \label{1scenario}
	\end{equation}
	or
	\begin{equation}
	 \frac{m}{M_p}<|Q|<\frac{M_p}{M_s}. \label{2scenario}
	\end{equation}
	The most interesting case is the one described by the second inequality, Eq. \eqref{2scenario}, which says that in {\it ghost-free} IDG we {\it can} study the gravitational field of an object whose charge is bigger than its mass (in units of Planck mass), allowing us to include, for example, an electron in IDG\footnote{ Note that for an elementary particle, like an electron, the Compton wavelength is much larger than its Schwarzschild radius, so in such a regime, quantum effects are not negligible and a classical theory of gravity alone cannot account for its gravitational effects. }.
	
	\item Finally, note that in the short-distance regime, $r<2/M_s$, one can in principle also have repulsive gravity, when $$Q^2M_s>m,$$ and it is only compatible with the scenario in Eq. \eqref{2scenario}, and not with Eq. \eqref{1scenario}, as can be easily checked. However, such a repulsive contribution would be confined within the region of non-locality, while outside, gravity would be still attractive as in the standard GR.

\end{itemize}

\section{Curvature tensors}\label{curv-tens}

We have computed all curvature tensors for the linearized IDG metric in Eqs. (\ref{eq:5},\ref{eq:15}). All curvature expressions are evaluated up to first order in $G$, as we are working in the linear regime ($h_{\mu \nu} \sim G$). The full expressions are shown in Appendix \ref{appendix C}; instead, in this section we wish to emphasize that the presence of non-local gravitational interaction is such that all the curvatures tensors and invariants are regularized at the origin, i.e. at $r=0$. Therefore, we will focus on their forms in the region of non-locality, $r<2/M_s$.

In the non-local dominated regime, many components of the curvature tensors vanish, while the remaining tend to {\it finite} constant values at the origin. The non-zero components of the Riemann tensor are given by 
\begin{equation}
\mathcal{R}_{0101} \sim \frac{1}{12} G M_s^3 \left(\frac{2 m}{\sqrt{\pi }}-Q^2 M_s\right); \\
\end{equation}
the non-zero components of the Ricci tensor:
\begin{equation}
\begin{array}{rl}
\mathcal{R}_{00} \sim & \displaystyle \frac{1}{4} G M_s^3 \left(\frac{2 m}{\sqrt{\pi }}-Q^2 M_s\right), \\
\mathcal{R}_{11} \sim & \displaystyle \frac{1}{12} G M_s^3 \left(\frac{6 m}{\sqrt{\pi }}-Q^2 M_s\right);
\end{array}
\end{equation}
and the Ricci scalar:
\begin{equation}
\mathcal{R} \sim \frac{G m M_s^3}{\sqrt{\pi }}.
\end{equation}
Finally, we note that all components of the Weyl tensor tend to zero at the origin, as we take $M_sr\rightarrow 0$:
\begin{equation}
\mathcal{C}_{\mu\nu\rho\sigma}\sim 0,
\end{equation}
Important point to note that with decreasing distance, different components of the Weyl tensor approach zero with different rates, particularly $O(r^2)$, $O(r^4)$ and $O(r^6)$ for small $r$, such that very close to the origin, we can imagine a constant valley, and the metric becomes exactly conformally-flat at $r=0$. This implies that the static metric of a charged source approaches {\it conformal-flatness} in the UV regime, $r<2/M_s.$  In fact, in the short distance regime, the metric in Eqs. (\ref{eq:5},\ref{eq:15}) can be approximated to:
\begin{equation}
ds^2\approx -(1-2A)dt^2+(1+2B)(dr^2+r^2d\Omega^2),
\end{equation}
which can be put in a conformally-flat form by introducing a {\it conformal time}, $\tau$, through the following coordinate transformation
\begin{equation}
\tau = \sqrt{\frac{1-2A}{1+2B}} t;
\end{equation}
such that the metric in the non-local region reads
\begin{equation}
\begin{array}{rl}
ds^2 = & (1+2B) \left[ -d\tau^2 + dr^2 + r^2 d\Omega^2 \right]\\
= & F^2\eta,
\end{array}
\end{equation}
where $\eta$ is the Minkowski metric and $F^2 \equiv 1+2B>0$ is the conformal factor. In Fig. 2 we have plotted the component $\mathcal{C}_{0101}$ of the Weyl tensor for both the charged case in {\it ghost-free} IDG and Reissner-Nordstr\"om in GR.

We can note that, by setting $Q=0$ we recover the curvature tensors for the {\it ghost-free} IDG neutral case obtained in Ref. \cite{Buoninfante:2018xiw}, as expected.
Furthermore, we have computed all the curvature invariants squared. Their full expressions are shown in Appendix \ref{appendix C}, while in this section we will present their values in the non-local region. In the non-local regime, $r<2/M_s,$ the Kretschmann scalar $\mathcal{K} = \mathcal{R}_{\mu \nu \rho \sigma} \mathcal{R}^{\mu \nu \rho \sigma}$ tends to the following finite constant value as $M_s r \rightarrow 0$:
\begin{equation}
\mathcal{K} \sim \frac{G^2 M_s^6 \left(10 m^2-6 \sqrt{\pi } m Q^2 M_s+\pi  Q^4 M_s^2\right)}{6 \pi }.
\end{equation}
In Fig. 3 we see explicitly that the Kretschmann invariant for a charged source in {\it ghost-free} IDG is regularized at the origin, unlike the case of Reissner Nordstr\"om in GR where we have a curvature singularity.
\begin{figure}[t]
	\centering
	\subfloat[Subfigure 1 list of figures text][Weyl component $\mathcal{C}_{0101}$]{
		\includegraphics[scale=0.44]{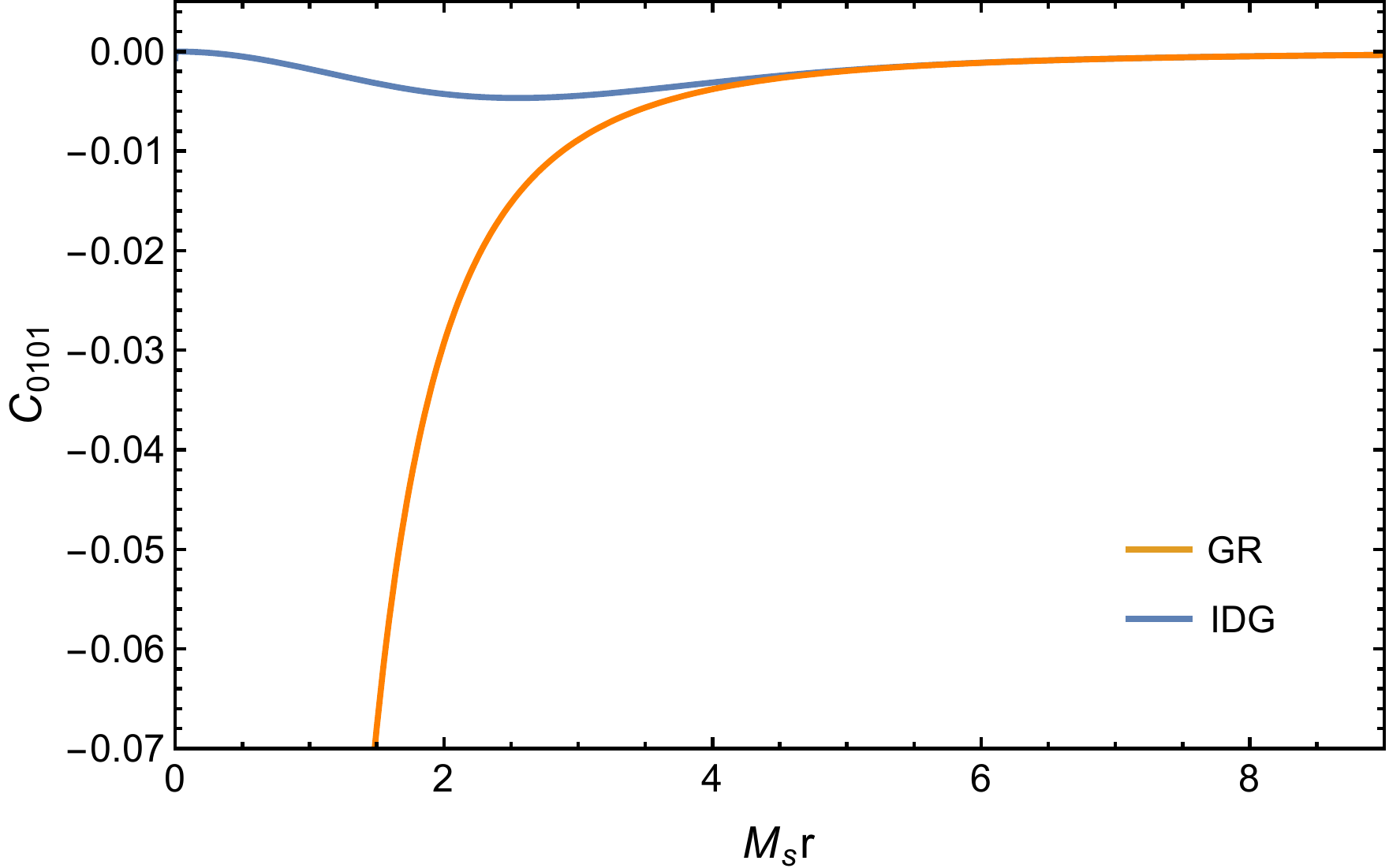}
		\label{fig.2.a}}
	\qquad
	\subfloat[Subfigure 2 list of figures text][Kretschmann invariant $\mathcal{K}$]{
		\includegraphics[scale=0.44]{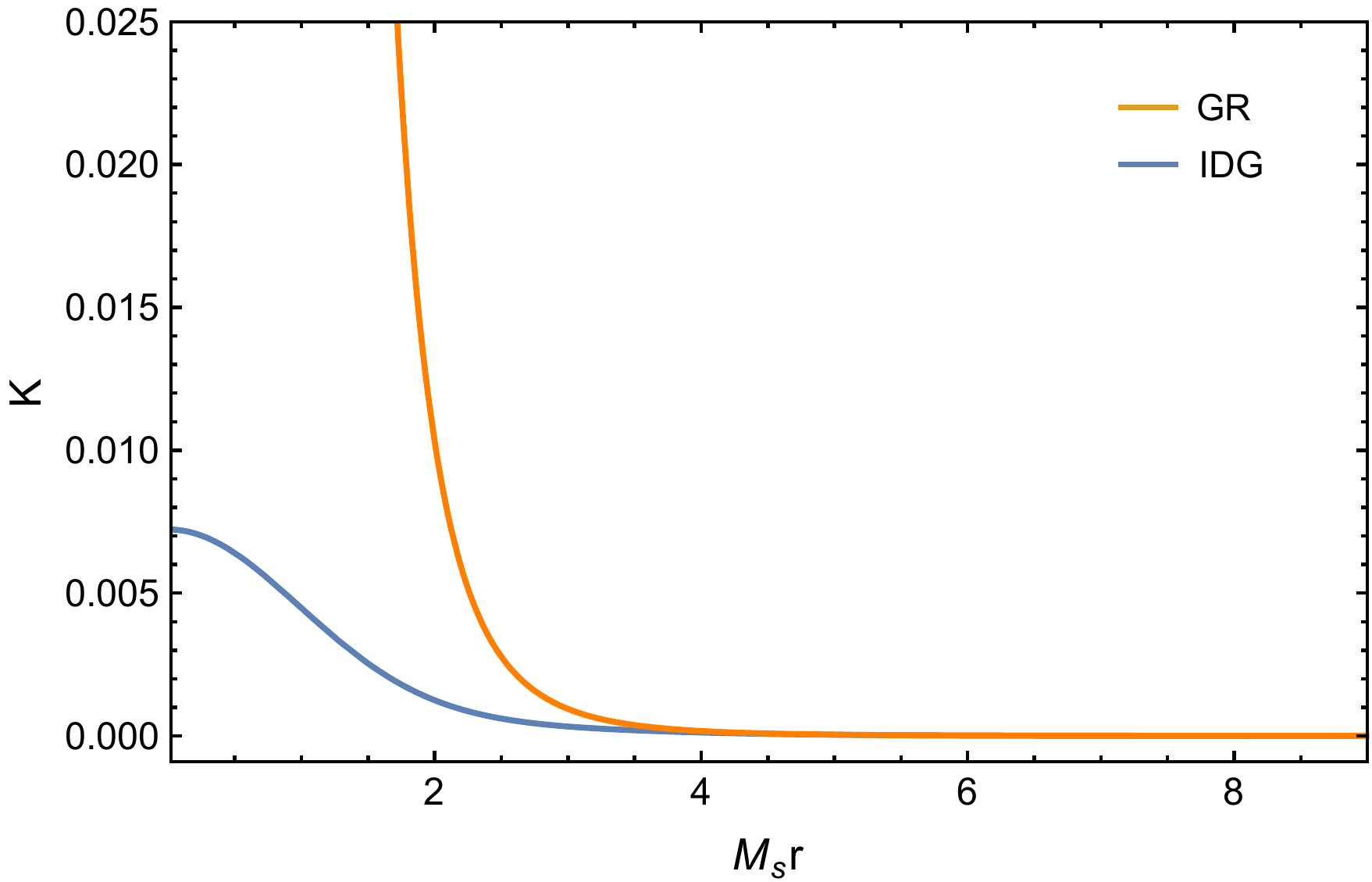}
		\label{fig.2.b}}
	\protect\caption{We have shown the spacetime properties of  {\it ghost-free} IDG (blue line) and GR (orange line) as functions of $M_sr$: (a) $\mathcal{C}_{0101}$ component of the Weyl tensor and (b) the Kretschmann scalar $\mathcal{K}$. We have set $M_p=1,~m=1,$ $M_s=0.5$ and $Q=0.5$. In these units the Schwarzschild radius is $r_{sch}=2$. Note that, unlike in GR where the linearized approximation would break down for $r\leq2$, in the case of {\it ghost-free} IDG we can smoothly approach $r=0$, provided $mM_s<M_p^2$ and $|Q|M_s<M_p.$ From these two plots it is obvious that the presence of non-locality helps to avoid the curvature-singularities at the origin.}\label{fig3}
\end{figure}

Moreover, in the non-local region, as $M_s r \rightarrow 0$, the non-zero components of the Ricci scalar squared and the Ricci tensor squared are given by the following finite constant values:
\begin{equation}
\mathcal{R}^2 \sim \frac{G^2 m^2 M_s^6}{\pi },
\end{equation}
\begin{equation}
\mathcal{R}_{\mu \nu} \mathcal{R}^{\mu \nu} \sim \frac{G^2 M_s^6 \left(12 m^2-6 \sqrt{\pi } m Q^2 M_s+\pi  Q^4 M_s^2\right)}{12 \pi }.
\end{equation}
While the Weyl tensor squared vanishes at the origin:
\begin{equation}
\mathcal{C}_{\mu \nu \rho \sigma} \mathcal{C}^{\mu \nu \rho \sigma} \sim 0.
\end{equation}

For completeness, it can be checked that the squared curvatures satisfy the following identity:
\begin{equation}
\mathcal{C}_{\mu \nu \rho \sigma} \mathcal{C}^{\mu \nu \rho \sigma} = \frac{\mathcal{R}^2}{3} - 2 \mathcal{R}_{\mu \nu} \mathcal{R}^{\mu \nu} + \mathcal{K}.
\end{equation}

Needless to say, all the curvature invariants for the case of a point-charge source in {\it ghost-free} IDG reduce to those of the uncharged case, $Q=0$, obtained in Ref. \cite{Buoninfante:2018xiw}, as expected.

\section{Conclusions}\label{concl}

In this paper we have found a linearized static metric solution for an electric point-charge in {\it ghost-free} IDG whose action is governed by Ref.\cite{Biswas:2011ar}. We have noticed that all the features of the case of neutral source are still kept; indeed the metric turns out to be regular at the origin and all curvature invariants are singularity-free. Moreover, all components of the Weyl tensor tend to zero as $M_s r \rightarrow 0$, meaning that the spacetime metric approaches conformal-flatness in the non-local regime.

We have also argued that in {\it ghost-free} IDG there is no need to require the cosmic censorship conjecture as the spacetime is singularity-free and devoid of any horizon as long as the inequalities in Eqs. (\ref{1-inequality},\ref{2-inequality}) hold true. Moreover, unlike the case of Reissner-Nordstr\"om in GR, in {\it ghost-free} IDG  {\it no} constraint has to be imposed on the relation between mass and charge of the source. Indeed, we can have cases where $|Q|>m/M_p$ is satisfied, allowing us to include in the theory objects such as an electron. Furthermore, we would expect that massive compact charged systems, devoid of singularity and event horizons can be found in this case too, similar to the case of neutral source studied in Ref.\cite{Koshelev:2017bxd}.

\acknowledgements The authors would like to thank Alan S. Cornell and Gaetano Lambiase for discussions. GEH is supported in part by the National Research Foundation of South Africa.

\appendix

\section{Reissner-Nordstr\"om metric in Einstein's general relativity}\label{Reiss-Nord}

In this appendix we wish to briefly review the main aspects of the Reissner-Nordstr\"om metric in GR; in particular we will introduce its form in the isotropic coordinates which is usually not studied in the standard text-books. The Reissner-Nordstr\"om metric is a non-vacuum solution of the full non-linear Einstein equations:
\begin{equation}
\mathcal{R}_{\mu\nu}-\frac{1}{2}g_{\mu\nu}\mathcal{R}=8\pi G\tau_{\mu\nu},\label{ap:1}
\end{equation}
where $\tau_{\mu\nu}$ is the electro-magnetic energy momentum tensor
\begin{equation}
\tau_{\mu\nu}=\frac{1}{4\pi}\left(\eta_{\rho\nu}F_{\mu\sigma}F^{\rho\sigma}-\frac{1}{4}\eta_{\mu\nu}F_{\rho\sigma}F^{\rho\sigma}\right).\label{ap:2}
\end{equation}
Assuming that no magnetic monopole is present, the only components of the field strength $F_{\mu\nu}$ will be given by the electric field part; then, due to spherical symmetry there will be a radial dependence so that the only non-vanishing components are $F_{10}=-F_{01}=Q/r^2.$ In the Schwarzschild coordinates $(t,R,\theta,\varphi)$, where $R$ is the polar radial coordinate, the metric reads
\begin{equation}
\begin{array}{rl}
ds^2= & \displaystyle -\left(1-\frac{2Gm}{R}+\frac{GQ^2}{R^2}\right)dt^2  \\
& \displaystyle +\left(1-\frac{2Gm}{R}+\frac{GQ^2}{R^2}\right)^{-1}dR^2+R^2d\Omega^2,
\end{array}\label{ap:3}
\end{equation}
where $d\Omega^2\equiv d\theta^2+{\rm sin}^2\theta d\varphi^2$, $m$ and $Q$ are the mass and the charge of the source, respectively.

One can immediately notice that the Reissner-Nordstr\"om metric posses {\it two} horizons that are defined as:
\begin{equation}
r_{\pm}=Gm\pm\sqrt{G^2m^2-GQ^2},\label{ap:4}
\end{equation}
where $r_+$ is called {\it outer} horizon, while $r_{-}$ {\it inner} horizon. 

We are now interested in finding the corresponding weak-field approximation for the metric in Eq. \eqref{ap:3} and to do so, as it happens for the Schwarzschild metric, it is more convenient to use the {\it isotropic coordinates} $(t,r,\theta,\varphi)$, where $r$ is the isotropic radial coordinate that in the case of Reissner-Nordstr\"om metric is defined by the following transformation:
\begin{equation}
R=r\left[\left(1+\frac{Gm}{2r}\right)^2-\frac{GQ^2}{4r^2}\right].\label{ap:5}
\end{equation}
In isotropic coordinates the metric in Eq. \eqref{ap:3} becomes:
\begin{equation}
\begin{array}{rl}
ds^2= & \displaystyle -\left[\frac{1-\left(\frac{Gm}{2r}\right)^2+\frac{GQ^2}{4r^2}}{\left(1+\frac{Gm}{2r}\right)^2-\frac{GQ^2}{4r^2}}\right]^2dt^2  \\
& \displaystyle +\left[\left(1+\frac{Gm}{2r}\right)-\frac{GQ^2}{4r^2}\right]^{2}(dr^2+r^2d\Omega^2);
\end{array}\label{ap:6}
\end{equation}
it is evident that for $Q=0$ we recover the Schwarzschild metric in isotropic coordinates. 

We can now make an expansion for weak gravitational field and stop at the linear order in $G$, thus the metric in Eq. \eqref{ap:6} in the linearized regime reads:
\begin{equation}
\begin{array}{rl}
ds^2= & \displaystyle -\left(1-\frac{2Gm}{r}+\frac{GQ^2}{r^2}\right)dt^2  \\
& \displaystyle +\left(1+\frac{2Gm}{r}-\frac{GQ^2}{2r^2}\right)(dr^2+r^2d\Omega^2).
\end{array}\label{ap:7}
\end{equation}

It is very important to stress that the metric in Eq. \eqref{ap:7} can be also obtained with a different procedure, i.e. by perturbing the metric around a Minkowski background and working in the linear regime:
\begin{equation}
g_{\mu\nu}(x)=\eta_{\mu\nu}+h_{\mu\nu}(x),\label{ap:8}
\end{equation}
so that the Einstein field equations in Eq. \eqref{ap:1} in the linear regime are given by:
\begin{equation}
\begin{array}{ll}
 \displaystyle \Box h_{\mu\nu}+(\eta_{\mu\nu}\partial_{\rho}\partial_{\sigma}h^{\rho\sigma}+\partial_{\mu}\partial_{\nu}h) & \\
 \,\,\,\,\,\,\,\,\,\,\,\displaystyle -\partial_{\sigma}(\partial_{\nu} h^{\sigma}_{\mu}+\partial_{\mu}h_{\nu}^{\sigma})-\eta_{\mu\nu}\Box h=-16\pi G\tau_{\mu\nu}. &
\end{array}\label{ap:9}
\end{equation}
By working in a weak field regime and considering the case of a static source, the spacetime metric can be written as:
\begin{equation}
ds^2=-(1+2\Phi(r))dt^2+(1-2\Psi(r))(dr^2+r^2d\Omega^2),\label{ap:10}
\end{equation}
where $r$ is the isotropic radial coordinate; thus the only two unknowns we need to find are the two gravitational potentials $\Phi$ and $\Psi$. By looking at the $00$-component and the trace of the linearized field equations in Eq. \eqref{ap:9} we obtain:
\begin{equation}
\begin{array}{rl}
\displaystyle \nabla^2h_{00}-\partial_i\partial_jh^{ij}+\nabla^2h=&-16\pi G\tau_{00},  \\
\displaystyle -2\nabla^2h+2\partial_i\partial_jh^{ij}=&-16\pi G\tau.
\end{array}\label{ap:11}
\end{equation}
Using $h=2(\Phi-3\Psi),$ $h_{00}=-2\Phi$ and $h_{ii}=-2\Psi$ we obtain two differential equations for the two gravitational fields $\Phi$ and $\Psi$:
\begin{equation}
\begin{array}{rl}
\displaystyle \nabla^2\Phi=&4\pi G(\tau+2\tau_{00}),  \\
\displaystyle \nabla^2\Psi=&4\pi G\tau_{00}.
\end{array}\label{ap:12}
\end{equation}
In the case of a charged static source, the energy-momentum tensor turns out to be traceless, $\tau=0,$ while the $00$-component is given by $\tau_{00}=Q^2/8\pi r^4.$ We can now solve the two differential equations in Eq. \eqref{ap:12} and obtain:
\begin{equation}
\begin{array}{rl}
\displaystyle \Phi(r)=&\displaystyle-\frac{C_1}{r}+\frac{GQ^2}{2r^2}+C_2,  \\
\displaystyle \Psi(r)=&\displaystyle-\frac{C_1}{r}+\frac{GQ^2}{4r^2}+C_2,
\end{array}\label{ap:13}
\end{equation}
where $C_1$ and $C_2$ are two integration constants whose value can be fixed by imposing physical suitable boundary conditions. First of all, for $r=\infty$ we want that $\Phi$ and $\Psi$ are zero (asymptotic flatness) which means $C_2=0$; while $C_1$ can be found by imposing that for $Q=0$ we have the usual Newtonian potential, thus $C_1=Gm:$
\begin{equation}
\begin{array}{rl}
\displaystyle \Phi(r)=&- \displaystyle \frac{Gm}{r}+\frac{GQ^2}{2r^2}, \\
\displaystyle \Psi(r)=&-\displaystyle \frac{Gm}{r}+\frac{GQ^2}{4r^2}.
\end{array}\label{ap:14}
\end{equation}
Notice that in the case of a charged source $\Phi\neq\Psi$. It is clear that the linearized metric in Eq. \eqref{ap:10} with the gravitational fields derived in Eq. \eqref{ap:14} coincides with the metric in Eq. \eqref{ap:7} which we derived instead by starting from the full non-linear Reissner-Nordstr\"om metric. It is very important to note that in the weak field approximation one has $2|\Phi|,2|\Psi|<1,$ which means that the radial coordinate has to always be greater than the outer horizon radius, $r>r_+$. Thus, in Einstein's GR it so happens that the linear approximation breaks down once one approaches the horizon. As we have seen in the paper, this is not the case for infinite derivative gravity, where (for both charged and neutral sources) the linear regime holds all the way up to $r=0.$

\section{Working with infinite order differential equations}\label{general-sol}

We can solve the differential equations in Eq. \eqref{eq:8} in several ways, for example by making the following temporary field-redefinitions:
\begin{equation}
\bar{\Phi}:=e^{-\nabla^2/M_s^2}\Phi,\,\,\,\,\,\,\,\,\,\,\,\bar{\Psi}:=e^{-\nabla^2/M_s^2}\Psi,\label{eq:9}
\end{equation}
so that the equations in \eqref{ap:8} become
\begin{equation}
\begin{array}{rl}
\displaystyle \nabla^2\bar{\Phi}(r)=&\displaystyle\frac{GQ^2}{r^4}  \\
\displaystyle \nabla^2\bar{\Psi}(r)=&\displaystyle\frac{GQ^2}{2r^4}.
\end{array}\label{eq:10}
\end{equation}
The structure of these equations is similar to the one we have in the case of a charged source in Einstein's GR, i.e. Reissner-Nordstr\"om, as also shown in Appendix \ref{Reiss-Nord} in Eq. \eqref{ap:12}; thus the solution will be also of the same structure as the one in Eq. \eqref{ap:13}:
\begin{equation}
\begin{array}{rl}
\displaystyle \bar{\Phi}(r)=&\displaystyle-\frac{C_1}{r}+\frac{GQ^2}{2r^2}+C_2  \\
\displaystyle \bar{\Psi}(r)=&\displaystyle-\frac{C_1}{r}+\frac{GQ^2}{4r^2}+C_2.
\end{array}\label{eq:11}
\end{equation}
By requiring the boundary conditions of asymptotic flatness one gets $C_2=0;$ while $C_1$ can be fixed that for $Q=0$ one recovers the case of neutral point-source, which means $C_1=Gm$. In terms of the real fields $\Phi$ and $\Psi$, Eq. \eqref{eq:9} will give:
\begin{equation}
\begin{array}{rl}
\displaystyle \Phi(r)=&\displaystyle-Gme^{\nabla^2/M_s^2}\left(\frac{1}{r}\right)+\frac{GQ^2}{2}e^{\nabla^2/M_s^2}\left(\frac{1}{r^2}\right)  \\
\displaystyle \Psi(r)=&\displaystyle-Gme^{\nabla^2/M_s^2}\left(\frac{1}{r}\right)+\frac{GQ^2}{4}e^{\nabla^2/M_s^2}\left(\frac{1}{r^2}\right)
\end{array}\label{eq:12}
\end{equation}
We now need to compute the action of the exponential $e^{-\nabla^2/M_s^2}$ on the functions $1/r$ and $1/r^2$. Notice that\footnote{We can also consider other choices of entire functions, for example we can generalize Eq. \eqref{choice} to $a(\Box_s)=e^{(-\Box_s)^n}$. In such a general case the factor $e^{{-k^2}/{M_s^2}}$ in the third equality of Eqs. (\ref{eq:13},\ref{eq:14}) is replaced by $e^{{-k^{2n}}/{M_s^{2n}}}$. For instance, for $n=2$, the potentials evaluate to generalized hypergeometric functions, and as expected, have non-singular constant values in the UV and recover GR in the IR.}:
\begin{equation}
\begin{array}{rl}
\displaystyle  e^{\nabla^2/M_s^2}\left(\frac{1}{r}\right)=& \displaystyle e^{\nabla^2/M_s^2}\int \frac{d^3k}{(2\pi)^3}\frac{4\pi}{k^2}e^{i\vec{k}\cdot\vec{r}}  \\
=& \displaystyle\int \frac{d^3k}{(2\pi)^3}\frac{4\pi}{k^2}e^{-k^2/M_s^2}e^{i\vec{k}\cdot\vec{r}}\\
=& \displaystyle\frac{2}{\pi}\int_{0}^{\infty} dk \ \frac{\sin{(kr)}}{kr} e^{-k^2/M_s^2}\\
=& \displaystyle\frac{1}{r}{\rm Erf}\left(\frac{M_s r}{2}\right),
\end{array}\label{eq:13}
\end{equation}
where we have used the fact that $4\pi/k^2$ is the Fourier transform of $1/r$. We can proceed in the same way to compute the second contribution, indeed by noting that the Fourier transform of $1/r^2$ is $\frac{2\pi^2}{k}{\rm sign}(k)$, one has:
\begin{equation}
\begin{array}{rl}
\displaystyle e^{\nabla^2/M_s^2}\left(\frac{1}{r^2}\right)=& \displaystyle e^{\nabla^2/M_s^2}\int \frac{d^3k}{(2\pi)^3}\frac{2\pi^2}{k}{\rm sign}(k)e^{i\vec{k}\cdot\vec{r}}  \\
=& \displaystyle  \int \frac{d^3k}{(2\pi)^3}\frac{2\pi^2}{k}{\rm sign}(k)e^{-k^2/M_s^2}e^{i\vec{k}\cdot\vec{r}}\\
=& \displaystyle\int_{0}^{\infty} dk \ \frac{\sin{(kr)}}{r} e^{-k^2/M_s^2}\\
=& \displaystyle \frac{M_s}{r}{\rm F}\left(\frac{M_s r}{2}\right),\label{eq:14}
\end{array}
\end{equation}
where ${\rm F}(M_{s}r/2)$ is the Dawson function.

Thus, by using Eqs. (\ref{eq:13},\ref{eq:14}), we can now obtain the expressions for the two metric potentials in Eq. \eqref{eq:15}.

\section{Full expressions of the curvature tensors}\label{appendix C}
In this appendix we show the full expressions for all curvature tensors and invariants for the IDG linearized static metric for a charged point-source derived in Eqs. (\ref{eq:5},\ref{eq:15}).

The Ricci scalar is given by:
\begin{equation}
\mathcal{R} = \frac{G m M_s^3 e^{-\frac{1}{4} r^2 M_s^2}}{\sqrt{\pi }};
\end{equation}
\begin{widetext}
the non-zero components of the Riemann tensor:	
\begin{equation}
\begin{array}{rl}
\mathcal{R}_{0101}=& \displaystyle \frac{1}{8} G \Bigg[M_s^3 \left(\frac{2 Q^2 \text{F}\left(\frac{r M_s}{2}\right)}{r}+\frac{4 m
   e^{-\frac{1}{4} r^2 M_s^2}}{\sqrt{\pi }}\right)+\frac{8 M_s \left(Q^2 \text{F}\left(\frac{r
   M_s}{2}\right)+\frac{2 m r e^{-\frac{1}{4} r^2 M_s^2}}{\sqrt{\pi }}\right)}{r^3}   \\
   & +\displaystyle Q^2 r M_s^5 \text{F}\left(\frac{r M_s}{2}\right)- \frac{16 m \text{Erf}\left(\frac{r
   M_s}{2}\right)}{r^3}-\frac{4 Q^2 M_s^2}{r^2}-Q^2 M_s^4\Bigg], \\
\mathcal{R}_{0303} =& \displaystyle \mathcal{R}_{0202} \sin^2 (\theta)  \\
			       =& \displaystyle \frac{1}{4} G \sin ^2(\theta ) \left[M_s \left(-\frac{2 Q^2 \text{F}\left(\frac{r
   M_s}{2}\right)}{r}-\frac{4 m e^{-\frac{1}{4} r^2 M_s^2}}{\sqrt{\pi }}\right)-Q^2 r
   M_s^3 \text{F}\left(\frac{r M_s}{2}\right)+\frac{4 m \text{Erf}\left(\frac{r
   M_s}{2}\right)}{r}+Q^2 M_s^2\right], \\
\mathcal{R}_{1313} =& \displaystyle \mathcal{R}_{1212} \sin^2 (\theta) \\
                                =& \displaystyle \frac{1}{16} G \sin ^2(\theta ) \Bigg[M_s \left(\frac{4 Q^2 \text{F}\left(\frac{r
   M_s}{2}\right)}{r}+\frac{16 m e^{-\frac{1}{4} r^2 M_s^2}}{\sqrt{\pi }}\right)+Q^2 r^3
   M_s^5 \text{F}\left(\frac{r M_s}{2}\right)-\frac{16 m \text{Erf}\left(\frac{r
   M_s}{2}\right)}{r} \\
   & +\displaystyle \frac{8 m r^2 M_s^3 e^{-\frac{1}{4} r^2 M_s^2}}{\sqrt{\pi }}-Q^2
   r^2 M_s^4-2 Q^2 M_s^2\Bigg], \\
\mathcal{R}_{2323} =& \displaystyle \frac{1}{4} G r \sin ^2(\theta ) \Bigg[M_s \left(-2 Q^2 \text{F}\left(\frac{r
   M_s}{2}\right)-\frac{8 m r e^{-\frac{1}{4} r^2 M_s^2}}{\sqrt{\pi }}\right)-Q^2 r^2
   M_s^3 \text{F}\left(\frac{r M_s}{2}\right)  \\
   & +\displaystyle 8 m \text{Erf}\left(\frac{r M_s}{2}\right)+Q^2 r
   M_s^2\Bigg];
  \end{array} 
\end{equation}
the non-zero components of the Ricci tensor:
\begin{equation}
\begin{array}{rl}
\mathcal{R}_{00} =& \displaystyle \frac{1}{8} G M_s^3 \left[\frac{Q^2 \left(r^2 M_s^2-2\right) \text{F}\left(\frac{r
   M_s}{2}\right)}{r}+\frac{4 m e^{-\frac{1}{4} r^2 M_s^2}}{\sqrt{\pi }} - Q^2 M_s\right],  \\
\mathcal{R}_{11} =& \displaystyle \frac{1}{4} G M_s \left[M_s \left(\frac{2 m M_s e^{-\frac{1}{4} r^2 M_s^2}}{\sqrt{\pi
   }}+\frac{Q^2}{r^2}\right)-\frac{Q^2 \left(r^2 M_s^2+2\right) \text{F}\left(\frac{r
   M_s}{2}\right)}{r^3}\right],  \\
\mathcal{R}_{22} =& \displaystyle \frac{1}{16} G M_s \left[\frac{Q^2 \left(r^4 M_s^4+4\right) \text{F}\left(\frac{r
   M_s}{2}\right)}{r}+M_s \left(\frac{8 m r^2 M_s e^{-\frac{1}{4} r^2 M_s^2}}{\sqrt{\pi
   }} - Q^2 r^2 M_s^2 - 2 Q^2\right)\right], \\
\mathcal{R}_{33} =& \displaystyle \frac{1}{16} G \sin ^2(\theta ) M_s \left[\frac{Q^2 \left(r^4 M_s^4+4\right)
   \text{F}\left(\frac{r M_s}{2}\right)}{r}+M_s \left(\frac{8 m r^2 M_s e^{-\frac{1}{4} r^2
   M_s^2}}{\sqrt{\pi }} - Q^2 r^2 M_s^2-2 Q^2\right)\right];
\end{array}
\end{equation}
the non-zero components of the Weyl tensor:
\begin{equation}
\begin{array}{rl}
\mathcal{C}_{0101} =& \displaystyle \frac{1}{48} G \Bigg[M_s^3 \left(\frac{12 Q^2 \text{F}\left(\frac{r M_s}{2}\right)}{r}+\frac{16
   m e^{-\frac{1}{4} r^2 M_s^2}}{\sqrt{\pi }}\right)+\frac{M_s \left(36 Q^2
   \text{F}\left(\frac{r M_s}{2}\right)+\frac{96 m r e^{-\frac{1}{4} r^2 M_s^2}}{\sqrt{\pi
   }}\right)}{r^3}   \\
   & +\displaystyle  3 Q^2 r M_s^5 \text{F}\left(\frac{r M_s}{2}\right)-\frac{96 m
   \text{Erf}\left(\frac{r M_s}{2}\right)}{r^3}-\frac{18 Q^2 M_s^2}{r^2}-3 Q^2
   M_s^4\Bigg], \\
\mathcal{C}_{0303} =& - \displaystyle \mathcal{C}_{1313} = \mathcal{C}_{0202} \sin^2 (\theta) = - \mathcal{C}_{1212} \sin^2 (\theta) \\
	                        =& \displaystyle \frac{1}{96} G \sin ^2(\theta ) \Bigg[4 r M_s^3 \left(-3 Q^2 \text{F}\left(\frac{r
   M_s}{2}\right)-\frac{4 m r e^{-\frac{1}{4} r^2 M_s^2}}{\sqrt{\pi }}\right)+M_s
   \left(-\frac{36 Q^2 \text{F}\left(\frac{r M_s}{2}\right)}{r}-\frac{96 m e^{-\frac{1}{4} r^2
   M_s^2}}{\sqrt{\pi }}\right) \\
   &-\displaystyle 3 Q^2 r^3 M_s^5 \text{F}\left(\frac{r M_s}{2}\right)+\frac{96 m
   \text{Erf}\left(\frac{r M_s}{2}\right)}{r}+3 Q^2 r^2 M_s^4+18 Q^2 M_s^2\Bigg] , \\
\mathcal{C}_{2323} =&\displaystyle \frac{1}{48} G r \sin ^2(\theta ) \Bigg[4 r^2 M_s^3 \left(-3 Q^2 \text{F}\left(\frac{r
   M_s}{2}\right)-\frac{4 m r e^{-\frac{1}{4} r^2 M_s^2}}{\sqrt{\pi }}\right)+M_s
   \left(-36 Q^2 \text{F}\left(\frac{r M_s}{2}\right)-\frac{96 m r e^{-\frac{1}{4} r^2
   M_s^2}}{\sqrt{\pi }}\right)\\
   &-\displaystyle 3 Q^2 r^4 M_s^5 \text{F}\left(\frac{r M_s}{2}\right)+96 m
   \text{Erf}\left(\frac{r M_s}{2}\right)+3 Q^2 r^3 M_s^4+18 Q^2 r M_s^2\Bigg].
\end{array}
\end{equation}
We will now show the expressions for the curvature invariants. The Ricci scalar squared is given by:
\begin{equation}
\mathcal{R}^2 = \displaystyle \frac{G^2 m^2 M_s^6 e^{-\frac{1}{2} r^2 M_s^2}}{\pi}; \\
\end{equation}
the Ricci tensor squared:
\begin{equation}
\begin{array}{rl}
\mathcal{R}_{\mu \nu} \mathcal{R}^{\mu \nu} =& \displaystyle \frac{G^2 M_s^2}{128 \pi  r^6} \Bigg[\text{F}\left(\frac{r M_s}{2}\right) \Bigg(32 \sqrt{\pi } m Q^2 r^7 M_s^6
   e^{-\frac{1}{4} r^2 M_s^2}-64 \sqrt{\pi } m Q^2 r^5 M_s^4 e^{-\frac{1}{4} r^2
   M_s^2} \\
   & \displaystyle -6 \pi  Q^4 r^7 M_s^7+4 \pi  Q^4 r^5 M_s^5-24 \pi  Q^4 r^3 M_s^3-48 \pi  Q^4 r
   M_s\Bigg)  \\
   & \displaystyle +\pi  Q^4 \left(3 r^8 M_s^8-8 r^6 M_s^6+24 r^4 M_s^4+32 r^2 M_s^2+48\right)
   \text{F}\left(\frac{r M_s}{2}\right)^2\\
   & +\displaystyle  M_s^4 \left(128 m^2 r^6 e^{-\frac{1}{2} r^2
   M_s^2}+4 \pi  Q^4 r^4\right)-32 \sqrt{\pi } m Q^2 r^6 M_s^5 e^{-\frac{1}{4} r^2
   M_s^2}+3 \pi  Q^4 r^6 M_s^6+12 \pi  Q^4 r^2 M_s^2\Bigg]; 
\end{array}
\end{equation}
the Weyl tensor squared:
\begin{equation}
\begin{array}{rl}
\mathcal{C}_{\mu \nu \rho \sigma} \mathcal{C}^{\mu \nu \rho \sigma} =& \displaystyle \frac{G^2 e^{-\frac{1}{2} r^2 M_s^2}}{192 \pi  r^6} \Bigg[-4 r^2 M_s^3 \left(3 \sqrt{\pi } Q^2
   e^{\frac{1}{4} r^2 M_s^2} \text{F}\left(\frac{r M_s}{2}\right)+4 m r\right) \\
   & -\displaystyle 12 M_s \left(3
   \sqrt{\pi } Q^2 e^{\frac{1}{4} r^2 M_s^2} \text{F}\left(\frac{r M_s}{2}\right)+8 m
   r\right)  \\
   & -\displaystyle 3 \sqrt{\pi } Q^2 r^4 M_s^5 e^{\frac{1}{4} r^2 M_s^2} \text{F}\left(\frac{r
   M_s}{2}\right)+96 \sqrt{\pi } m e^{\frac{1}{4} r^2 M_s^2} \text{Erf}\left(\frac{r
   M_s}{2}\right)  \\
   & +\displaystyle 18 \sqrt{\pi } Q^2 r M_s^2 e^{\frac{1}{4} r^2 M_s^2}+3 \sqrt{\pi } Q^2
   r^3 M_s^4 e^{\frac{1}{4} r^2 M_s^2}\Bigg]^2;
\end{array}
\end{equation}
and the Kretschmann invariant:
\begin{equation}
\begin{array}{rl}
\mathcal{K} =& \displaystyle \frac{e^{-\frac{1}{2} r^2 M_s^2} G^2}{32 \pi  r^6} \Bigg[3 e^{\frac{1}{2} r^2 M_s^2} \pi  Q^4 r^8
   {\rm F}\left(\frac{r M_s}{2}\right){}^2 M_s^{10}-6 e^{\frac{1}{2} r^2 M_s^2} \pi  Q^4 r^7
   {\rm F}\left(\frac{r M_s}{2}\right) M_s^9 \\
   & +\displaystyle \left(32 e^{\frac{1}{4} r^2 M_s^2} m \sqrt{\pi }
   Q^2 {\rm F}\left(\frac{r M_s}{2}\right) r^7+3 e^{\frac{1}{2} r^2 M_s^2} \pi  Q^4 r^6+8
   e^{\frac{1}{2} r^2 M_s^2} \pi  Q^4 {\rm F}\left(\frac{r M_s}{2}\right){}^2 r^6\right)
   M_s^8  \\
   & -\displaystyle 4 e^{\frac{1}{4} r^2 M_s^2} \sqrt{\pi } Q^2 r^5 \left(7 e^{\frac{1}{4} r^2
   M_s^2} \sqrt{\pi } {\rm F}\left(\frac{r M_s}{2}\right) Q^2+8 m r\right) M_s^7  \\
   & +\displaystyle 4 r^4 \left(18e^{\frac{1}{2} r^2 M_s^2} \pi  {\rm F}\left(\frac{r M_s}{2}\right){}^2 Q^4+5 e^{\frac{1}{2}
   r^2 M_s^2} \pi  Q^4+32 e^{\frac{1}{4} r^2 M_s^2} m \sqrt{\pi } r {\rm F}\left(\frac{r
   M_s}{2}\right) Q^2+24 m^2 r^2\right) M_s^6  \\
   & -\displaystyle 24 e^{\frac{1}{4} r^2 M_s^2} \sqrt{\pi }
   Q^2 r^3 \left(8 m r+e^{\frac{1}{4} r^2 M_s^2} \sqrt{\pi } {\rm F}\left(\frac{r
   M_s}{2}\right) \left(5 Q^2+4 m r \text{Erf}\left(\frac{r M_s}{2}\right)\right)\right)
   M_s^5  \\
   & +\displaystyle 4 r^2 \bigg(40 e^{\frac{1}{2} r^2 M_s^2} \pi  {\rm F}\left(\frac{r M_s}{2}\right){}^2
   Q^4+15 e^{\frac{1}{2} r^2 M_s^2} \pi  Q^4+144 e^{\frac{1}{4} r^2 M_s^2} m \sqrt{\pi }
   r {\rm F}\left(\frac{r M_s}{2}\right) Q^2  \\
   & \displaystyle +24 e^{\frac{1}{2} r^2 M_s^2} m \pi  r
   \text{Erf}\left(\frac{r M_s}{2}\right) Q^2+ 128 m^2 r^2\bigg) M_s^4  \\
   & \displaystyle -16 e^{\frac{1}{4}
   	r^2 M_s^2} \sqrt{\pi } r \bigg(3 e^{\frac{1}{4} r^2 M_s^2} \sqrt{\pi } {\rm F}\left(\frac{r
   	M_s}{2}\right) \left(5 Q^2+8 m r \text{Erf}\left(\frac{r M_s}{2}\right)\right) Q^2\\
   & +\displaystyle 4 mr \left(9 Q^2+8 m r \text{Erf}\left(\frac{r M_s}{2}\right)\right)\bigg) M_s^3+48
   \bigg(5 e^{\frac{1}{2} r^2 M_s^2} \pi  {\rm F}\left(\frac{r M_s}{2}\right){}^2 Q^4\\
   & \displaystyle +24
   e^{\frac{1}{4} r^2 M_s^2} m \sqrt{\pi } r {\rm F}\left(\frac{r M_s}{2}\right) Q^2  +\displaystyle 4 m r \left(3 e^{\frac{1}{2} r^2 M_s^2} \pi  \text{Erf}\left(\frac{r M_s}{2}\right) Q^2+8 m
   r\right)\bigg) M_s^2 \\
   & -\displaystyle 384 e^{\frac{1}{4} r^2 M_s^2} m \sqrt{\pi } \left(3
   e^{\frac{1}{4} r^2 M_s^2} \sqrt{\pi } {\rm F}\left(\frac{r M_s}{2}\right) Q^2+8 m r\right)
   \text{Erf}\left(\frac{r M_s}{2}\right) M_s+1536 e^{\frac{1}{2} r^2 M_s^2} m^2 \pi 
   \text{Erf}\left(\frac{r M_s}{2}\right)^2\Bigg].
\end{array}
\end{equation}
\end{widetext}
In the case $Q=0$, we would recover all curvature tensors and invariants for the case of a neutral point-source obtained in Ref. \cite{Buoninfante:2018xiw}, as expected.

\end{document}